\documentclass[showpacs,nofootinbib,onecolumn,11pt, beta 3),aps,floats,showkeys]{revtex4}
\usepackage{amsmath}
\usepackage{graphicx}
\usepackage{dcolumn}
\usepackage{bm}

\input{tcilatex}

\begin{document}

\preprint{}
\title{Noise Correlation Induced Synchronization in a Mutualism Ecosystem}
\author{Wei-Rong Zhong}
\email{wr-zhong@126.com}
\author{Yuan-Zhi Shao}
\author{Zhen-Hui He}
\author{Meng-Jie Bie}
\author{Dan Huang}
\altaffiliation[ ]{Corresponding Author}
\affiliation{State Key Laboratory of Optoelectronic Materials and Technologies,\\
School of Physics \& Engineering, Sun Yat-sen University, Guangzhou 510275,
People's Republic of China}

\begin{abstract}
Understanding the cause of the synchronization of population evolution is an
important issue for ecological improvement. Here we present a
Lotka-Volterra-type model driven by two correlated environmental noises and
show, via theoretical analysis and direct simulation, that noise correlation
can induce a synchronization of the mutualists. The time series of mutual
species exhibit a chaotic-like fluctuation, which is independent to the
noise correlation, however, the chaotic fluctuation of mutual species ratio
decreases with the noise correlation. A quantitative parameter defined for
characterizing chaotic fluctuation provides a good approach to measure when
the complete synchronization happens.
\end{abstract}

\maketitle

Synchronization is a common phenomenon in nature (Pikovsky et al. 2001),
e.g., in physical systems (Pecora \& Carroll 1990), chemical systems (Neiman
et al. 1999) and biosystems (Winfree 1990). Specially, the population
synchrony has become an important issue in population biology (Benton et al.
2001). Additionally, throughout the twentieth century, a dominant study has
focused on the stochastic fluctuation in population abundance (Cushing et
al. 1998; Begon et al. 1996). Interestingly, synchronization and
stochasticity regarded as two independent phenomena attract more and more
concerns (Higgins et al. 1997; Leirs et al. 1997; Grenfell et al. 1998;
Keeling 2000; Tuljapurkar \& Haridas 2006). Recently, Benton et al. (2001)
suggested that environmental noise plays a determined role in the population
synchrony.

Regrettably, due to the difficulties in mathematical analyses and
experimental demonstrations, the study of relationship between noise
correlation and population synchrony is still insufficient. However,
surprising phenomena emerge sometimes just because there exists correlation
or collaboration; one of the most well known examples is the self-organized
behavior (Bak et al. 1988). We know that two stochastic processes in
ecosystems maybe originate from the same source like nature disasters and
epidemics. For example, epidemics can lead one kind of species in a
mutualism system as well as another to death synchronously, thus it is
reasonable to suppose that the population fluctuations of the two species of
mutualism system are correlated.

Mutualism is an association between two species that benefits both. For
example, Van der Heijden et al. (1998) have argued that mycorrhizal fungi
determine plant species diversity. As the number of fungal species per
system was increased, so the collective biomass of shoots and roots
increased, as did the diversity of plant community. Namely the number of two
species has a positive linear relationship, indicating a synchronous
fluctuation. For this reason, we focus on the Lotka-Volterra-type
two-species model with correlated stochastic components. The noise
correlation we will consider here induces some novel synchronous phenomena
in a mutualism ecosystem that are not found before. Our main methods will be
expected to have applicability to other ecosystems, since they have been
also of similar correlations of two populations.

Additionally, noise-induced chaos is observed in many experimental and
theoretical situations (Miller and Greeberg 1992; Kovanis et al. 1995; Ruiz
1995; Gao et al. 1999; Jing \& Yang 2006). Can noise induce chaos in a
mutualism ecosystem? How does this happen? When is the chaos
synchronization? In this paper, we will give physical explanations for these
problems.

Mutualistic relationships can be modeled with equations similar to the
Lotka-Volterra competition equations.%
\begin{eqnarray}
\frac{dx}{dt} &=&r_{1}x-k_{1}x^{2}+\gamma _{1}xy, \\
\frac{dy}{dt} &=&r_{2}y-k_{2}y^{2}+\gamma _{2}xy,
\end{eqnarray}%
where $\gamma _{1}$ and $\gamma _{2}$ are the positive effects of species $y$
on species $x$ and of species $x$ on species $y$, respectively. $r_{1}$ and $%
r_{2}$ are respectively growth rate of species $x$ and that of species $y$. $%
k_{1}$, $k_{2}$ are the parameters related to carrying capacities while one
of the mutualists is alone. Due to environmental noise like epidemics,
natural disasters and adventitious species, realistic mutualism systems do
not follow the solutions of the above deterministic differential equations
exactly. The simplest way to include these environmental noises is to add
stochastic terms to the above deterministic differential equations and the
equivalent stochastic differential equations are%
\begin{eqnarray}
\frac{dx}{dt} &=&r_{1}x-k_{1}x^{2}+\gamma _{1}xy+x\xi (t), \\
\frac{dy}{dt} &=&r_{2}y-k_{2}y^{2}+\gamma _{2}xy+y\eta (t),
\end{eqnarray}%
where $\xi (t)$ and $\eta (t)$ are referred to as the birth rate
fluctuations of species $x$ and those of species $y$, respectively. Consider
the synchronization of the two fluctuations, we define them as correlated
noises. They are Gaussian white noises satisfying%
\begin{eqnarray}
\langle \xi (t)\xi (t^{\prime })\rangle &=&2M_{x}\delta (t-t^{\prime }), \\
\langle \eta (t)\eta (t^{\prime })\rangle &=&2M_{y}\delta (t-t^{\prime }), \\
\langle \xi (t)\eta (t^{\prime })\rangle &=&2\lambda \sqrt{M_{x}M_{y}}\delta
(t-t^{\prime }),
\end{eqnarray}%
in which $M_{x},M_{y}$ are the intensities of noises; $\lambda $, ranges
from zero to one, denotes the correlation coefficient between $\xi (t)$ and $%
\eta (t)$, i.e., the correlation degree between species $x$ and species $y$.
When $\lambda \ $equal zero, the growth rate fluctuations of species $x$ and
those of species $y$ are independent. When $\lambda $ equal $1$, their
growth rate fluctuations are completely correlated. $\delta (t-t^{\prime })$
is a Dirac delta function under different moments.

In the absence of noises, it can be shown that the system has a stable
equilibrium state ($x_{0}$, $y_{0}$) when $t\rightarrow \infty $ for $\gamma
_{1}\gamma _{2}<k_{1}k_{2}$, where $x_{0}=(k_{2}r_{1}+\gamma
_{1}r_{2})/(k_{1}k_{2}-\gamma _{1}\gamma _{2})$ and $y_{0}=$ $%
(k_{1}r_{2}+\gamma _{2}r_{1})/(k_{1}k_{2}-\gamma _{1}\gamma _{2})$,
describes a limit point of system (Stiling 2002). For the sake of
simplicity, we set \textit{k}$_{1}$\textit{=k}$_{2}$\textit{=r}$_{1}$\textit{%
=r}$_{2}$\textit{=1} and $\gamma _{1}$\textit{=}$\gamma _{2}$\textit{=0.5}.
Figure 1 shows that the system can lead to a stable point where the
isoclines cross. In the presence of noises, we define quantitative
parameters, i.e., the mean population densities of the mutualists, which are
useful to display the influences of noises on the mutualism ecosystem. The
mean population densities of the mutualists are respectively written as%
\begin{eqnarray}
\langle x\rangle &=&\int_{0}^{\infty }\int_{0}^{\infty }xp(x,y)dxdy, \\
\langle y\rangle &=&\int_{0}^{\infty }\int_{0}^{\infty }yp(x,y)dxdy,
\end{eqnarray}%
and their variances are $\sigma _{x}^{2}=\langle x^{2}\rangle -\langle
x\rangle ^{2}$ and $\sigma _{y}^{2}=\langle y^{2}\rangle -\langle y\rangle
^{2}$, respectively. Here $p(x,y)$ is the joint stationary probability
distribution of species $x$ and $y$ satisfying $\int_{0}^{\infty
}\int_{0}^{\infty }p(x,y)dxdy=1.$

\ifcase\msipdfoutput
\FRAME{ftbpFU}{3.1851in}{2.5719in}{0pt}{\Qcb{Graphic model of the mutualism
ecosystem is based on modified Lotka-Volterra equations for \textit{k}$_{1}$%
\textit{=k}$_{2}$\textit{=r}$_{1}$\textit{=r}$_{2}$\textit{=1} and $\gamma
_{1}$\textit{=}$\gamma _{2}$\textit{=0.5.} There is a stable equilibrium
point.}}{}{fig1.eps}{%
\special{language "Scientific Word";type "GRAPHIC";maintain-aspect-ratio
TRUE;display "USEDEF";valid_file "F";width 3.1851in;height 2.5719in;depth
0pt;original-width 4.248in;original-height 3.4238in;cropleft "0";croptop
"1";cropright "1";cropbottom "0";filename '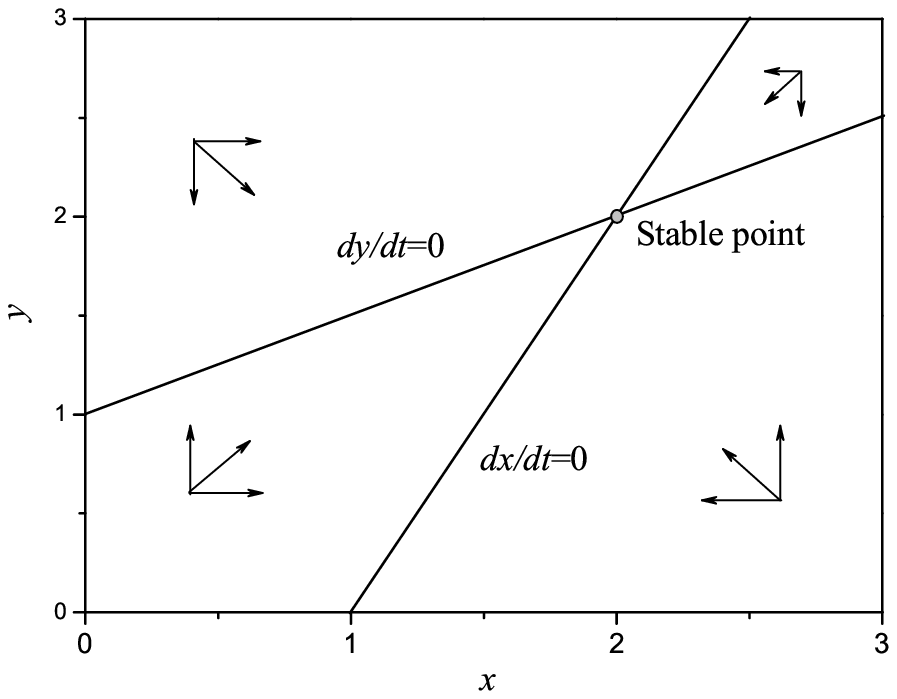';file-properties
"XNPEU";}}%
\else
\begin{figure}[ptb]\begin{center}
\includegraphics[
natheight=3.4238in, natwidth=4.248in, height=2.5719in, width=3.1851in]
{C:/Documents and Settings/LuckyStar/桌面/mutualism/graphics/Fig1__1.pdf}%
\caption{Graphic model of the mutualism ecosystem is based on modified
Lotka-Volterra equations for \textit{k}$_{1}$\textit{=k}$_{2}$\textit{=r}$%
_{1}$\textit{=r}$_{2}$\textit{=1} and $\protect\gamma _{1}$\textit{=}$%
\protect\gamma _{2}$\textit{=0.5.} There is a stable equilibrium point.}
\end{center}\end{figure}%
\fi

The equivalent stationary Fokker-Planck equations of Eqs.(3) and (4) can be
derived and the joint stationary probability distributions can be obtained
for two independent noises (Cai and Lin 2004). However, they are far
difficult to be solved for correlated noises, here we obtain the joint
stationary probability distributions of the population densities of the
mutualists via a direct simulation of Eqs.(3) and (4) (Zhong et al. 2006).

To describe noise-induced chaos and the relationship between chaos and
synchrony, we need to define chaos carefully. For example, given the time
series \{$x(t)$\}, an $m$-dimensional phase portrait is reconstructed with
delay coordinates, i.e., a point on the attractor is given by $%
\{x(t),x(t+\tau ),...,x(t+[m-1]\tau )\}$, with $m$ being the embedding
dimension and $\tau $ being the delay time (here we choose $m=2,$and $\tau
=2 $) (Wolf et al. 1985; Gao et al. 1999). We denote the initial distance
between these two points $L(t_{0})$. At a later time $t_{1}$, the initial
length will have evolved to length $L^{^{\prime }}(t_{1})$. Using Wolf
method, we can give lyapunov exponent as following%
\begin{equation}
\Lambda _{x}=\langle \frac{1}{t_{N}-t_{0}}\sum_{k=1}^{N}\ln \frac{%
L^{^{\prime }}(t_{k})}{L(t_{k-1})}\rangle
\end{equation}%
in which $N$ is total number of replacement steps.

It is comprehensible that noises can induce instability. Figure 2(a)
illustrates that the populations of species $x$ and those of species $y$
driven by environmental noise are instable. However, when the noise
correlation increases, shown in Figs.2(b) and 2(c), more and more
populations of species $x$ and those of species $y$ concentrate on a line in
which the ratio of species $y$ to species $x$ is constant. When the
correlation equals 1, the ratio of species $y$ to species $x$ is independent
on the noise, i.e., the population fluctuations of species $y$ and those of
species $x$ are synchronization. This extreme situation can be analyzed from
the Eqs.(3) and (4). If the noise correlation equals 1, $\xi (t)$ of
Eq.(3)can be regarded as same as $\eta (t)$ of Eq.(4) for the same noise
intensity. If considering $\frac{d\frac{y}{x}}{dt}=\frac{1}{x}\frac{dy}{dt}-%
\frac{y}{x^{2}}\frac{dx}{dt}$, and substituting Eqs.(3) and (4) for it, we
obtain%
\begin{equation}
\frac{d\frac{y}{x}}{dt}=(k_{1}+\gamma _{2})y-(k_{2}+\gamma _{1})\frac{y^{2}}{%
x}
\end{equation}%
Considered a steady solution $\frac{d\frac{y}{x}}{dt}=0$, then Eq(11) can be
written as $\frac{y}{x}=\frac{k_{1}+\gamma _{2}}{k_{2}+\gamma _{1}}=1$,
i.e., the populations of species $x$ have a positive linear relationship
with those of species $y$, shown in Fig.2(c).

\ifcase\msipdfoutput
\FRAME{itbpF}{3.3408in}{2.6264in}{0in}{}{}{fig2a.eps}{%
\special{language "Scientific Word";type "GRAPHIC";maintain-aspect-ratio
TRUE;display "USEDEF";valid_file "F";width 3.3408in;height 2.6264in;depth
0in;original-width 4.2756in;original-height 3.3529in;cropleft "0";croptop
"1";cropright "1";cropbottom "0";filename '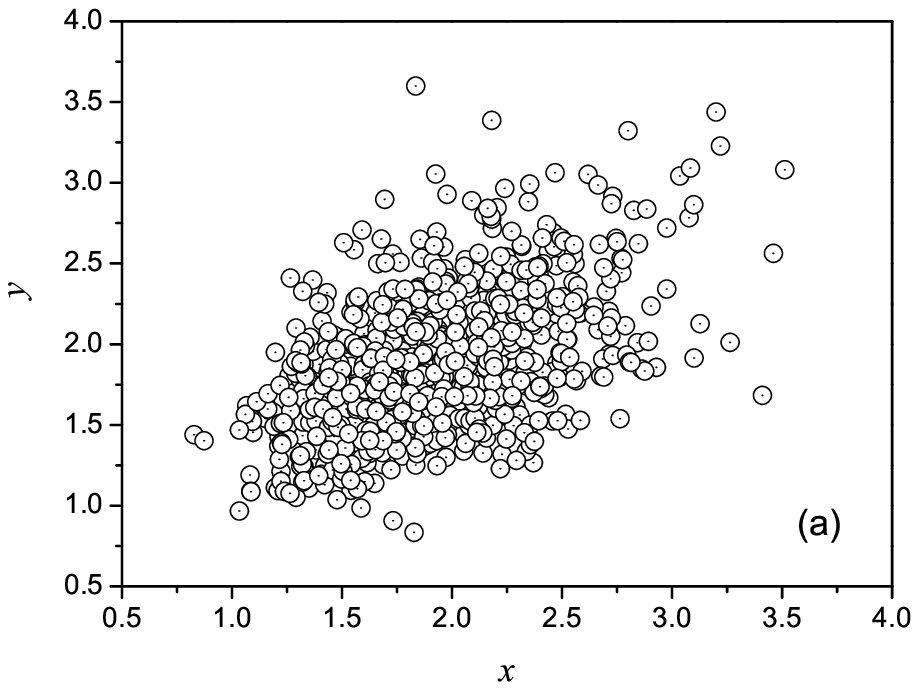';file-properties
"XNPEU";}}%
\else
\includegraphics[
natheight=3.3529in, natwidth=4.2756in, height=2.6264in, width=3.3408in]
{C:/Documents and Settings/LuckyStar/桌面/mutualism/graphics/Fig2a__2.pdf}%
\fi
\ifcase\msipdfoutput
\FRAME{itbpF}{3.3364in}{2.6152in}{0in}{}{}{fig2b.eps}{%
\special{language "Scientific Word";type "GRAPHIC";maintain-aspect-ratio
TRUE;display "USEDEF";valid_file "F";width 3.3364in;height 2.6152in;depth
0in;original-width 4.2886in;original-height 3.3529in;cropleft "0";croptop
"1";cropright "1";cropbottom "0";filename '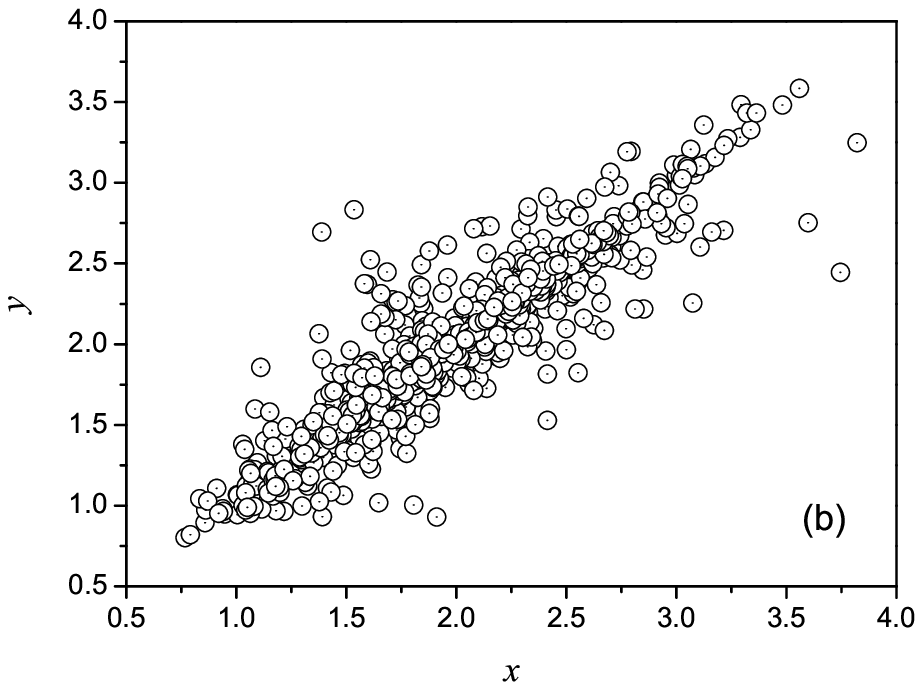';file-properties
"XNPEU";}}%
\else
\includegraphics[
natheight=3.3529in, natwidth=4.2886in, height=2.6152in, width=3.3364in]
{C:/Documents and Settings/LuckyStar/桌面/mutualism/graphics/Fig2b__3.pdf}%
\fi

\ifcase\msipdfoutput
\FRAME{ftbphFU}{3.3434in}{2.6204in}{0pt}{\Qcb{Distributions of the
mutualists driven by environmental noise with different correlations. (a) $%
\lambda =0.0$; (b) $\lambda =0.8$; (c) $\lambda =1.0,$ including simulated
results (circle) and theoretical analysis (solid line) results. The
parameter are $M_{x}=M_{y}=0.8.$}}{}{fig2c.eps}{%
\special{language "Scientific Word";type "GRAPHIC";maintain-aspect-ratio
TRUE;display "USEDEF";valid_file "F";width 3.3434in;height 2.6204in;depth
0pt;original-width 4.2886in;original-height 3.3529in;cropleft "0";croptop
"1";cropright "1";cropbottom "0";filename '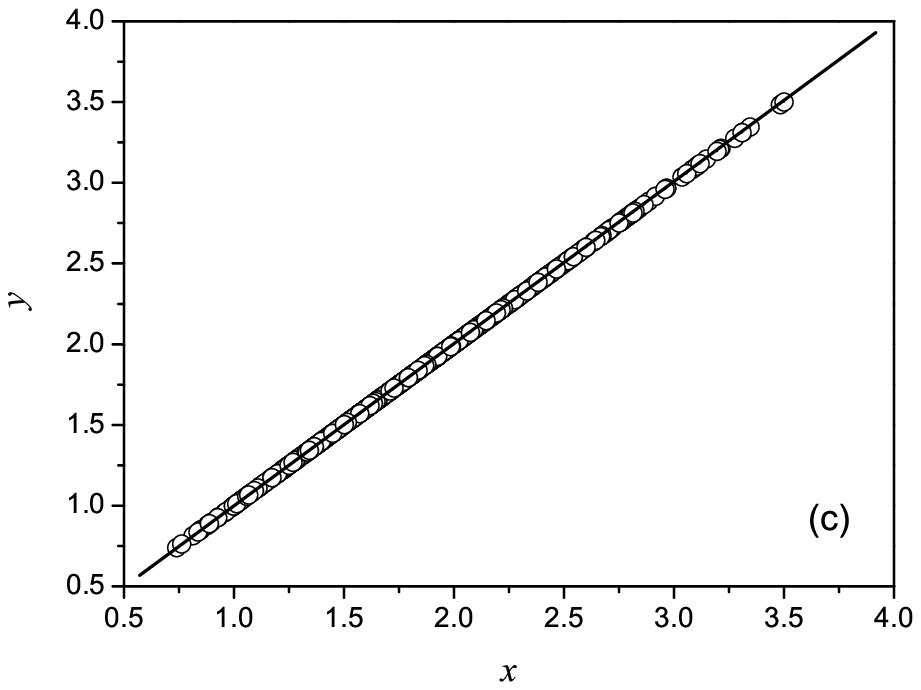';file-properties
"XNPEU";}}%
\else
\begin{figure}[hptb]\begin{center}
\includegraphics[
natheight=3.3529in, natwidth=4.2886in, height=2.6204in, width=3.3434in]
{C:/Documents and Settings/LuckyStar/桌面/mutualism/graphics/Fig2c__4.pdf}%
\caption{Distributions of the mutualists driven by environmental noise with
different correlations. (a) $\protect\lambda =0.0$; (b) $\protect\lambda =0.8
$; (c) $\protect\lambda =1.0,$ including simulated results (circle) and
theoretical analysis (solid line) results. The parameter are $%
M_{x}=M_{y}=0.8.$}
\end{center}\end{figure}%
\fi

The linear relationships between the variance of $\frac{y}{x}$ and the noise
correlation are given in Fig.3, indicating the variance of $\frac{y}{x}$
decreases with the noise correlation. When the correlation equals 1, the
fluctuations of the ratio of species $y$ to species $x$ are extinct.%
\ifcase\msipdfoutput
\FRAME{ftbpFU}{3.3408in}{2.5737in}{0pt}{\Qcb{Variances of the ratio of
species $y$ and species $x$ vs. noise correlations under different noise
intensities. The remaining parameter are the same as for Fig.2. }%
}{}{fig3.eps}{%
\special{language "Scientific Word";type "GRAPHIC";maintain-aspect-ratio
TRUE;display "USEDEF";valid_file "F";width 3.3408in;height 2.5737in;depth
0pt;original-width 4.401in;original-height 3.3814in;cropleft "0";croptop
"1";cropright "1";cropbottom "0";filename '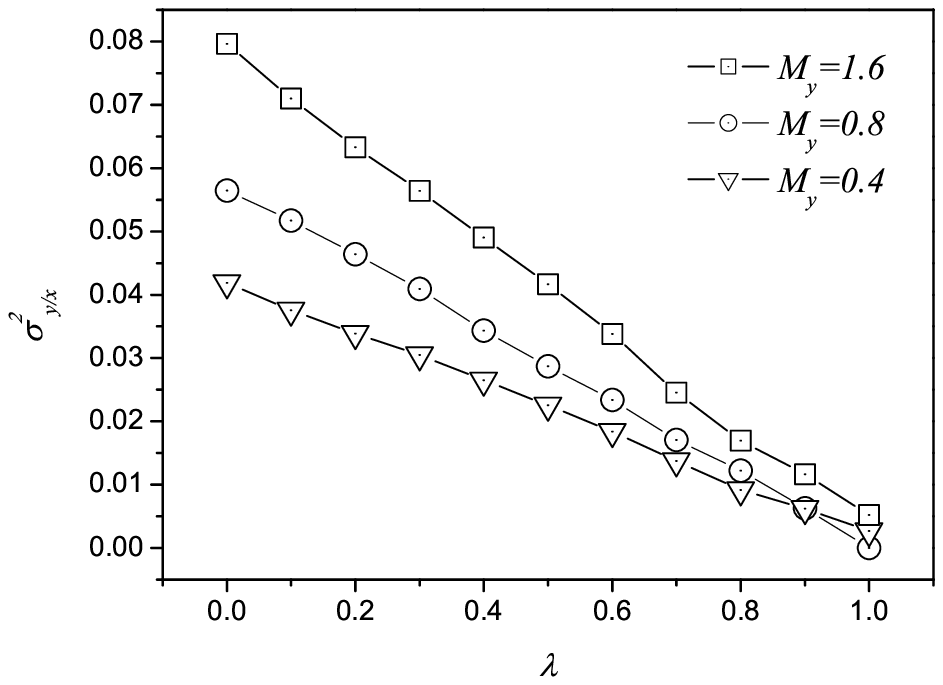';file-properties
"XNPEU";}}%
\else
\begin{figure}[ptb]\begin{center}
\includegraphics[
natheight=3.3814in, natwidth=4.401in, height=2.5737in, width=3.3408in]
{C:/Documents and Settings/LuckyStar/桌面/mutualism/graphics/Fig3__5.pdf}%
\caption{Variances of the ratio of species $y$ and species $x$ vs. noise
correlations under different noise intensities. The remaining parameter are
the same as for Fig.2. }
\end{center}\end{figure}%
\fi

Figure 4 shows that the lyapunov exponents of species $y$ and those of
species $x$ fluctuate between 0.2 and 0.35 when the noise correlation
increases, however, the lyapunov exponent of the ratio of species $y$ to
species $x$ drops to zero. Generally speaking, when lyapunov exponent is
positive, the time series are regarded as disorder or chaotic
characteristic. On the contrary, if lyapunov exponent is non-positive, the
time series are of order or periodic characteristic. Figure 4 indicates that
two disorder time series with correlation can exhibit an order behavior, and
the synchronization in a mutualism ecosystem happens at the lyapunov
exponent equals zero.%
\ifcase\msipdfoutput
\FRAME{ftbpFU}{3.3304in}{2.5728in}{0pt}{\Qcb{Lyapunov exponents of species $y
$, species $x$ and their ratio vary with noise correlations. The completely
synchronous fluctuation point corresponds to zero-lyapunov exponent.}%
}{}{fig4.eps}{%
\special{language "Scientific Word";type "GRAPHIC";maintain-aspect-ratio
TRUE;display "USEDEF";valid_file "F";width 3.3304in;height 2.5728in;depth
0pt;original-width 4.3163in;original-height 3.3252in;cropleft "0";croptop
"1";cropright "1";cropbottom "0";filename '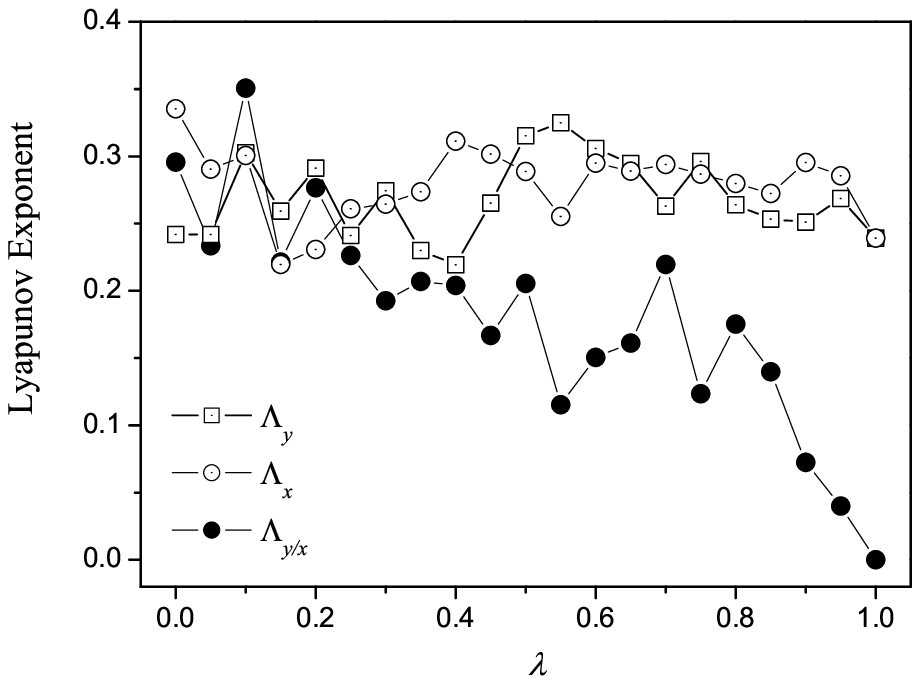';file-properties
"XNPEU";}}%
\else
\begin{figure}[ptb]\begin{center}
\includegraphics[
natheight=3.3252in, natwidth=4.3163in, height=2.5728in, width=3.3304in]
{C:/Documents and Settings/LuckyStar/桌面/mutualism/graphics/Fig4__6.pdf}%
\caption{Lyapunov exponents of species $y$, species $x$ and their ratio vary
with noise correlations. The completely synchronous fluctuation point
corresponds to zero-lyapunov exponent.}
\end{center}\end{figure}%
\fi

In summary, the environmental noise correlation can cause a synchronous
fluctuation in mutualism ecosystems, and this synchronous fluctuation can be
characterized by the lyapunov exponent. If environmental noise correlation
originates from the correlation degree of the mutualists, our main results
illustrate that the mutualism ecosystems with closer relationship have more
stable evolutions. This is a helpful illumination for ecologists to study
ecosystem balance.

This work was partially supported by the National Natural Science Foundation
(Grant No. 60471023) P. R. China.

Bak, P., Tang, C. \& Wiesenfeld, K. (1988). Self-organized criticality.
Phys. Rev. A, 38, 364-374.

Begon, M., Harper, J. \& Townsend, C. (1996). Ecology: Individuals,
populations and communities, 3rd edn. Blackwell Science, Oxford.

Benton, T.G., Lapsley, C.T., \& Beckerman, A.P. (2001). Population synchrony
and environmental variation: an experimental demonstration. Ecology Letter,
4, 236-243.

Cai, G.Q. \& Lin, Y.K. (2004). Stochastic analysis of the Lotka-Volterra
model for ecosystems. Phys. Rev. E, 70, 0419101-7.

Cushing, J. M., Costantino, R. F., Dennis, B., Desharnais, R. A. \& Henson,
S.M. (1998). Nonlinear population dynamics: Models, experiments and data. J.
Theoret. Biol., 194, 1-9.

Gao, J.B., Hwang, S.K. \& Liu, J.M. (1999). When can noise induce chaos?
Phys. Rev. Lett., 82, 1132-1135.

Grenfell, B.T., Wilson, K., Finkenstadt, B.F., Coulson, T.N., Murray, S.,
Albon, S.D., Pemberton, J.M., Clutton-Brock, T.H. \& Crawley, M.J. (1998).
Noise and determinism in synchronized sheep dynamics. Nature, 394, 674-677.

Higgins, K., Hastings, A., Sarvela, J.N. \& Botsford, L.W. (1997).
Stochastic dynamics and deterministic skeletons: Population behaviour of
Dungeness crab. Science, 276, 1431-1435.

Jing, Z.J. \& Yang, J.P. (2006). Bifurcation and chaos in discrete-time
predator--prey system. Chaos, Solitons \& Fractals, 27, 259-277.

Keeling, M. (2000). Metapopulation moments: coupling, stochasticity and
persistence. J. Anim. Ecol., 69, 725-736.

Kovanis, V., Gavrielides, A., Simpson, T. B. \& Liu J. M. (1995).
Instabilities and chaos in optically injected semiconductor lasers. Appl.
Phys. Lett., 67, 2780-2782.

Leirs, H., Stenseth, N.C., Nichols, J. D., Hines, J.E., Verhagen, R. \&
Verheyen, W. (1997). Stochastic seasonality and nonlinear density-dependent
factors regulate population size in an African rodent. Nature, 389, 176-180.

Miller, Paul A. \& Greenberg, K. E. (1992). Period-doubling bifurcation in a
plasma reactor. Appl. Phys. Lett., 60, 2859-2861.

Neiman, A., Schimansky-Geier, L. \& Cornell-Bell, A. (1999). Noise-enhanced
phase synchronization in excitable media. Phys. Rev. Lett., 83, 4896-4899.

Pecora, L.M. \& Carroll, T.L. (1990). Synchronization in chaotic system.
Phys. Rev. Lett., 64, 821-824.

Pikovsky, A., Rosenblum, M. \& Kurths, J. (2001). Synchronization: A
universal concept in nonlinear sciences. Cambridge University Press, New
York.

Ruiz, G.A. (1995). Period doubling bifurcations in cardiac systems. Chaos,
Solitons \& Fractals, 6, 487-494.

Stiling, P. (2002). Ecology: Theories and applications, 4th edn.
Prentice-Hall, New Jersey.

Tuljapurkar, S \& Haridas, C.V. (2006). Temporal autocorrelation and
stochastic population growth. Ecology Letter, 9, 327-337.

Van der Heijden, M. G. A., Klironomos, J. N., Ursic, M., Moutoglis, P.,
Streitwolf-Engel, R., Boller, T., Wiemken, A., \& Sanders, I. R. (1998).
Mycorrhizal fungal biodiversity determines plant biodiversity, ecosystem
variability and productivity. Nature, 396: 69-72.

Winfree, A.T. (1990). Geometry of biological time. Springer-Verlag, New York.

Wolf, A., Swift, J.B., Swinney, H.L. \& Vastano, J.A. (1985). Determining
lyapunov exponents from a time series. Physica D, 16, 285-317.

Zhong, W.R., Shao, Y.Z. \& He, Z.H. (2006). Correlated noises in a
prey-predator ecosystem. Chin. Phys. Lett., 23, 742-745.

\end{document}